\documentclass[12pt]{article}
\usepackage{feynarts}
\usepackage[dvips]{graphicx}
\usepackage[usenames,dvips]{color}
\usepackage{here}
\usepackage{mcite}
\usepackage{amsmath}
\usepackage{bbm}
\usepackage{a4wide}

\allowdisplaybreaks

\def\re{\mathop{\rm Re}}
\def\d{\mathrm{d}}

\def\be#1\ee{\begin{equation}#1\end{equation}}
\def\ba#1\ea{\begin{align}#1\end{align}}

\newcommand{\diana}{\textsc{Diana}}
\newcommand{\form}{\textsc{Form}}
\newcommand{\fortran}{\textsc{Fortran}}
\newcommand{\topfit}{\textsc{Topfit}}
\newcommand{\looptools}{\textit{LoopTools}}
\newcommand{\FA}{\textit{FeynArts}}
\newcommand{\FC}{\textit{FormCalc}}

\newcommand{\eq}[1]{Eq.~(\ref{#1})}

\newcommand{\fig}[1]{Fig.~\ref{#1}}
\newcommand{\tab}[1]{Tab.~\ref{#1}}

\renewcommand{\d}[1]{\mathrm{d}#1}

\renewcommand{\slash}[1]{/ \hspace{-6.5pt}#1}

\newcommand{\red}{\color{Red}}

\def \oa{$\mathcal{O}(\alpha)$~}
\def \Oa{\mathcal{O}(\alpha)}
\def \Oatwo{\mathcal{O}(\alpha^2)}
\def \Oathree{\mathcal{O}(\alpha^3)}
\def \Oafour{\mathcal{O}(\alpha^4)}
\def \Mcal{\mathcal{M}}
\newcommand \Mcalf[2]{\mathcal{M}^{(#1)}_{#2}}
\newcommand \Mcals[2]{{\mathcal{M}^{(#1)}_{#2}}^*}

\def \ct{\cos{\theta}}
\def \emo{\cdot{} 10^{-1}}
\def \emt{\cdot{} 10^{-2}}

\begin{document}

\title{
{\bf\oa electroweak corrections to the processes
\begin{boldmath}
$e^+ e^- \to \tau^-\tau^+, c\bar c, b\bar b, t\bar t$ 
\end{boldmath}
-- a comparison --}\footnote{%
  Work supported in part by the European Community's Human Potential
  Programme under contract HPRN--CT--2000--00149 ``Physics at Colliders''
   and by Sonderforschungsbereich/Transregio 9 of DFG
  ``Computergest\"utzte Theoretische Teilchenphysik.''
  This research has also been supported by a Marie Curie Fellowship of 
  the European Community's Research Training Project
  under contract HPMF--CT--2002--01694%
.}}

\author{{\sc T.~Hahn$^{a}$\footnote{%
    E-mails: alejandro.lorca@desy.de, tord.riemann@desy.de,
             anja.werthenbach@cern.ch, \hfill
             hahn@mppmu.mpg.de, hollik@mppmu.mpg.de.},
  ~~W.~Hollik$^{a}$,}
 \\
  {\sc A.~Lorca$^{b}$,
  ~~T.~Riemann$^{b}$, ~~and
  ~~A.~Werthenbach$^{c}$} \\
{}\\
{\small $^{a}${\em Max-Planck-Institut f\"ur Physik,
                   F\"ohringer Ring 6,
                   D--80805 Munich, Germany}}
\\
{\small $^{b}${\em DESY Zeuthen,
                   Platanenallee 6,
                   D--15738 Zeuthen, Germany}} 
\\
{\small $^{c}${\em CERN,
                   TH Division,
                   CH--1211 Gen\`eve 23, Switzerland}} 
\\
}

\date{}
\maketitle
\vspace*{-11cm}
\noindent
DESY 03--086          
\\  
MPP--2003--24
\\
SFB/CPP-03-13
\\
hep-ph/0307132
\\
\vspace*{8cm}
\begin{abstract}%
We present the electroweak one-loop corrections to the processes
$e^+e^-\to f\bar f$, $f = \tau, c, b, t$, at energies relevant for a
future linear collider.  
The results of two independent calculations are
compared and agreement is found at a technical-precision level of ten to
twelve digits. 
\end{abstract}

\section{Introduction}

With the advent of the next linear collider (LC), center-of-mass
energies will rise up to several hundred GeV and the envisioned
luminosity will be as high as 300 fb$^{-1}$.  Evidently, a new era of
precision physics is approaching.  The experimental precision which can
be achieved at such a machine will by far exceed all current standards
and will be a challenge to experimentalists and theoreticians alike.  To
obtain reliable predictions for the next generation of linear colliders,
the inclusion of electroweak one-loop corrections becomes essential.

Two-fermion production processes, such as
\be
e^+e^-\to f\bar f (\gamma)\,,
\label{eeffprocess}
\ee
play a leading role at typical LC energies as foreseen by
\cite{Aguilar-Saavedra:2001rg}.  In the late seventies the one-loop
correction to muon-pair production was calculated for the first time
\cite{Passarino:1979jh}, where the muons were considered to be massless.
Ever since, fermion-pair production processes attracted attention and
various masses were successively introduced into the calculation.
Recently, a high degree of computational precision was achieved in
numerically comparing various results on radiative corrections to
top-pair production (see \cite{Fleischer:2002rn,Fleischer:2002nn} and
references therein).  Such comparisons are invaluable to ensure the
establishment of reliable, well-tested codes.

Here, we extend the study \cite{Fleischer:2002rn} to other final states.
In this particular comparison we do not include hard bremsstrahlung. 
This issue has been discussed in detail in
\cite{Fleischer:2002nn,Fleischer:2003kk} and will be calculated for
realistic applications by dedicated Monte-Carlo programs for 2- to
6-fermion production
\cite{Kolodziej:2001xe,Dittmaier:2002ap,Dittmaier:2002zc}.


\section{Cross-section formulae}

\subsection{Notation and conventions}

In this section, we will outline the framework to compute electroweak
corrections to differential and total cross-sections in \oa of the
electromagnetic coupling.  This includes one-loop amplitudes as well as
soft-photon bremsstrahlung.

\begin{figure}[h]
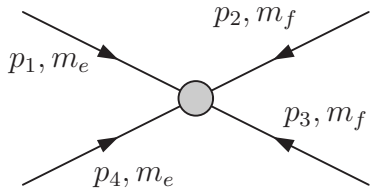

\begin{center}
\vspace*{-5ex}
\begin{feynartspicture}(144,144)(1,1)
\FADiagram{}
\FAProp(0.,15.)(10.,10.)(0.,){/Straight}{1}
\FALabel(3.89862,12.6172)[tr]{$p_1, m_e$}
\FAProp(0.,5.)(10.,10.)(0.,){/Straight}{1}
\FALabel(4.10138,6.01724)[tl]{$p_4, m_e$}
\FAProp(20.,15.)(10.,10.)(0.,){/Straight}{1}
\FALabel(15.8986,13.9828)[br]{$p_2, m_f$}
\FAProp(20.,5.)(10.,10.)(0.,){/Straight}{1}
\FALabel(15.1014,8.18276)[bl]{$p_3, m_f$}
\FAVert(10.,10.){-1}
\end{feynartspicture}
\vspace*{-7ex}
\end{center}
\caption{Definitions of the kinematical variables.}
\label{twototwo}
\end{figure}

In a $2\to 2$-particle process we follow the momenta and mass convention 
of \fig{twototwo}:
\be
\frac{\d\sigma}{\d\ct} = \frac 1{32\pi}
\frac{\beta_{f}}{s\beta_{e}} 
{\sum_{\mathrm{conf}} \left| \Mcal_{ef} \right|}^2 \,,
\label{dxscost}
\ee
where $\theta$ is the scattering angle. Furthermore we have
\ba
\beta_i &\equiv \sqrt{1 - 4\frac{m_i^2}{s}}\,
\\
s &\equiv (p_1 + p_4)^2 = E_{\text{CM}}^2\,
\label{sman} \\
t &\equiv (p_1 + p_2)^2 = -\frac s2 (1 - \beta_e\beta_f\ct) + m_e^2 + m_f^2\,
\label{tman} \\
u &\equiv (p_1 + p_3)^2 = -\frac s2 (1 + \beta_e\beta_f\ct) + m_e^2 + m_f^2\,.
\label{uman}
\ea


\subsection{Unpolarized cross-section}
\label{ua}

We consider only the unpolarized cross-section and thus have to
average over initial spin configurations ($\sigma_e$), sum over the
final ones ($\sigma_f$), and add incoherently the number of colours
($C_f$) which cannot be distinguished:
\be
\sum_{\mathrm{conf}} |\Mcal_{ef}|^2 =
\frac 14\sum_{\sigma_e=1}^4 \sum_{\sigma_f=1}^4 C_f |\Mcal_{ef}|^2\,.
\label{polarizationsum}
\ee
The invariant transition amplitude $\Mcal_{ef}$ can be expressed in
terms of a standard basis of matrix elements $M_i$, containing all the
kinematical information of the interaction, and the form factors $F_i$,
which account for the pure dynamical part:
\be
\Mcal_{ef}
=\sum_{i} M^{\phantom{|}}_{i} F_{i}\,.
\label{amplitude}
\ee


\subsection{Neglecting the electron mass}

In this comparative study we are neglecting the electron mass $m_e$ in
the purely weak contributions at the diagrammatic level, i.e.\ we
neglect diagrams containing the electron--Higgs Yukawa coupling, which
is proportional to the electron mass.  This simplifies the final
expression significantly and minimizes the number of independent form
factors.  We do not neglect the electron mass elsewhere so as to safely
compute the photonic corrections.


\subsection{Structure of \oa corrections}

The hierarchy of contributions in the perturbative expansion of the
$2\to2$ cross-section reads
\ba
|\Mcal|^2 &= |\Mcalf{0}{ef} + \Mcalf{1}{ef} + \ldots|^2 + 
             |\Mcalf{0}{_\gamma} + \ldots|^2 \nonumber \\
&= \underbrace{ \phantom{\Big|}\Mcals{0}{ef} \Mcalf{0}{ef} \phantom{\Big|} }_{\Oatwo} +
   \underbrace{2\re \Big( \Mcals{0}{ef} \Mcalf{1}{ef} \Big) +
     \Mcals{0}{\gamma} \Mcalf{0}{\gamma}}_{\Oathree} +
   \underbrace{\phantom{\Big|} \ldots \phantom{\Big|}}_{\Oafour}
\ea


Soft-photon contributions are added to remove the infrared singularities
of the photonic self-energies, vertices, and boxes.  

For the Born amplitude, an appropriate basis for the matrix elements is:
\begin{equation}
\begin{array}{l@{~\equiv~}lll@{~\otimes~}lll}
M_1&\bar{v}_e(p_4,\sigma_{e^+})&\gamma^\mu \mathbbm{1} &u_e(p_1,\sigma_{e^-})&
    \bar{u}_f(-p_2,\sigma_{f})&\gamma_\mu\mathbbm{1}&v_f(-p_3,\sigma_{\bar{f}})\\
M_2&\bar{v}_e(p_4,\sigma_{e^+})&\gamma^\mu \mathbbm{1} &u_e(p_1,\sigma_{e^-})&
    \bar{u}_f(-p_2,\sigma_{f})&\gamma_\mu \gamma_5 &v_f (-p_3,\sigma_{\bar{f}})\\
M_3&\bar{v}_e(p_4,\sigma_{e^+})&\gamma^\mu \gamma_5 &u_e(p_1,\sigma_{e^-})&
    \bar{u}_f(-p_2,\sigma_{f})&\gamma_\mu \mathbbm{1} &v_f(-p_3,\sigma_{\bar{f}})\\
M_4&\bar{v}_e(p_4,\sigma_{e^+})&\gamma^\mu \gamma_5 &u_e(p_1,\sigma_{e^-})&
    \bar{u}_f(-p_2,\sigma_{f})&\gamma_\mu \gamma_5 &v_f(-p_3,\sigma_{\bar{f}})\,.
\end{array}
\label{bornbasis}
\end{equation}
The differential Born cross-section finally reads
\ba
\frac{\d\sigma}{\d\ct}\bigg|_\mathrm{Born}
=& \, \frac 1{32\pi} \frac{\beta_f}{\beta_e} C_f \bigg\{ s(1+\beta_e^2\beta_f^2\cos^2\theta) 
   \left(|F^{(0)}_1|^2+|F^{(0)}_2|^2+|F^{(0)}_3|^2+|F^{(0)}_4|^2\right)\nonumber \\
&  + 2s\beta_e\beta_f\ct \left({F^{(0)}_1}^*F^{(0)}_4 + 
    {F^{(0)}_2}^*F^{(0)}_3 + {F^{(0)}_3}^*F^{(0)}_2 +
    {F^{(0)}_4}^*F^{(0)}_1\right) \nonumber \\
&  + 4 (m_f^2+m_e^2) \left(|F^{(0)}_1|^2 - |F^{(0)}_4|^2\right) +
4(m_f^2-m_e^2)\left(-|F^{(0)}_2|^2 + |F^{(0)}_3|^2\right)\nonumber\\
&  +16\frac{m_f^2m_e^2}{s}\left( -|F^{(0)}_2|^2-|F^{(0)}_3|^2+2|F^{(0)}_4|^2\right)
    \bigg\} \,,
\label{borndxs}
\ea
with the form factors
\ba
F^{(0)}_1 &= i e^2 \bigg(+V_e V_f \frac{1}{s-M_Z^2+iM_Z\Gamma_Z}+
  Q_eQ_f\frac{1}{s}  \bigg)
\label{f1} \\
F^{(0)}_2 &= i e^2 \bigg(-V_e A_f \frac{1}{s-M_Z^2+iM_Z\Gamma_Z} \bigg)
\label{f2} \\
F^{(0)}_3 &= i e^2 \bigg(-A_e V_f \frac{1}{s-M_Z^2+iM_Z\Gamma_Z} \bigg)
\label{f3} \\
F^{(0)}_4 &= i e^2 \bigg(+A_e A_f \frac{1}{s-M_Z^2+iM_Z\Gamma_Z} \bigg) \,.
\label{f4}
\ea

\begin{figure}[ht]
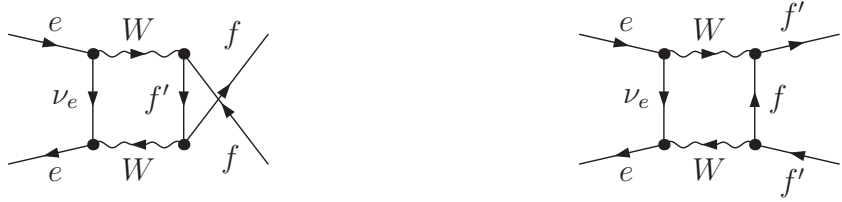

\begin{center}
\vspace*{-5ex}
\begin{feynartspicture}(324,108)(3,1)
\FADiagram{}
\FAProp(0.,15.)(6.5,13.5)(0.,){/Straight}{1}
\FALabel(3.59853,15.2803)[b]{$e$}
\FAProp(0.,5.)(6.5,6.5)(0.,){/Straight}{-1}
\FALabel(3.59853,4.71969)[t]{$e$}
\FAProp(20.,15.)(13.5,6.5)(0.,){/Straight}{-1}
\FALabel(17.9814,13.8219)[br]{$f$}
\FAProp(20.,5.)(13.5,13.5)(0.,){/Straight}{1}
\FALabel(17.8314,6.42814)[tr]{$f$}
\FAProp(6.5,13.5)(6.5,6.5)(0.,){/Straight}{1}
\FALabel(5.43,10.)[r]{$\nu_e$}
\FAProp(6.5,13.5)(13.5,13.5)(0.,){/Sine}{1}
\FALabel(10.,14.57)[b]{$W$}
\FAProp(6.5,6.5)(13.5,6.5)(0.,){/Sine}{-1}
\FALabel(10.,5.43)[t]{$W$}
\FAProp(13.5,6.5)(13.5,13.5)(0.,){/Straight}{-1}
\FALabel(12.43,10.)[r]{$f'$}
\FAVert(6.5,13.5){0}
\FAVert(6.5,6.5){0}
\FAVert(13.5,6.5){0}
\FAVert(13.5,13.5){0}

\FADiagram{}

\FADiagram{}
\FAProp(0.,15.)(6.5,13.5)(0.,){/Straight}{1}
\FALabel(3.59853,15.2803)[b]{$e$}
\FAProp(0.,5.)(6.5,6.5)(0.,){/Straight}{-1}
\FALabel(3.59853,4.71969)[t]{$e$}
\FAProp(20.,15.)(13.5,13.5)(0.,){/Straight}{-1}
\FALabel(16.4015,15.2803)[b]{$f'$}
\FAProp(20.,5.)(13.5,6.5)(0.,){/Straight}{1}
\FALabel(16.4015,4.71969)[t]{$f'$}
\FAProp(6.5,13.5)(6.5,6.5)(0.,){/Straight}{1}
\FALabel(5.43,10.)[r]{$\nu_e$}
\FAProp(6.5,13.5)(13.5,13.5)(0.,){/Sine}{1}
\FALabel(10.,14.57)[b]{$W$}
\FAProp(6.5,6.5)(13.5,6.5)(0.,){/Sine}{-1}
\FALabel(10.,5.43)[t]{$W$}
\FAProp(13.5,13.5)(13.5,6.5)(0.,){/Straight}{-1}
\FALabel(14.57,10.)[l]{$f$}
\FAVert(6.5,13.5){0}
\FAVert(6.5,6.5){0}
\FAVert(13.5,13.5){0}
\FAVert(13.5,6.5){0}

\end{feynartspicture}
\vspace*{-7ex}
\end{center}
\caption{Electroweak W--W-box diagrams at the one-loop level,
  where $f$ denotes an isospin-up and $f'$ an isospin-down fermion.}
\label{boxdiagrams}
\end{figure}

The one-loop calculations for the different fermion flavours are very
similar: Only the W--W-box diagram is different for different values of
the isospin of the final-state fermion (see \fig{boxdiagrams}).  These
weak box diagrams were suppressed in applications to LEP1 physics but
started to become numerically important at LEP2.  They were studied
systematically e.g.\ in Section 2.2 of \cite{Boudjema:1996qg} and
Section 5.4 of \cite{Kobel:2000aw}, but a comparison with the published
numbers is not straightforward. 

At the one-loop level, with the appearance of vertex and box diagrams,
the Lorentz structure of the matrix element is enriched:
\be
\begin{array}{l@{~=~}lrl@{~\otimes~}lrl}
M_{1,k}&\bar{v}_e(p_4,\sigma_{e^+})&\gamma^\mu \{\mathbbm{1},\gamma_5\} &u_e(p_1,\sigma_{e^-})&
    \bar{u}_f(-p_2,\sigma_{f})&\gamma_\mu \{\mathbbm{1},\gamma_5\} &v_f(-p_3,\sigma_{\bar{f}})\\
M_{2,k}&\bar{v}_e(p_4,\sigma_{e^+})& \slash{p}_2 \{\mathbbm{1},\gamma_5\} &u_e(p_1,\sigma_{e^-})&
    \bar{u}_f(-p_2,\sigma_{f})& \slash{p}_4 \{\mathbbm{1},\gamma_5\}
    &v_f(-p_3,\sigma_{\bar{f}})\\
M_{3,k}&\bar{v}_e(p_4,\sigma_{e^+})& \slash{p}_2 \{\mathbbm{1},\gamma_5\} &u_e(p_1,\sigma_{e^-})&
    \bar{u}_f(-p_2,\sigma_{f})& \{\mathbbm{1},\gamma_5\} &v_f (-p_3,\sigma_{\bar{f}})\\
M_{4,k}&\bar{v}_e(p_4,\sigma_{e^+})&\gamma^\mu \{\mathbbm{1},\gamma_5\} &u_e(p_1,\sigma_{e^-})&
    \bar{u}_f(-p_2,\sigma_{f})&\gamma_\mu \slash{p}_4
\{\mathbbm{1},\gamma_5\} &v_f(-p_3,\sigma_{\bar{f}}) \,,
\end{array}
\label{fullbasis}
\ee
where the index $k$ stands for the four possible combinations of
$\{\mathbbm{1},\gamma_5\}~\otimes~\{\mathbbm{1},\gamma_5\}$ as in
\eq{bornbasis}, leading to a basis of 16 elements.  The one-loop
contribution to the cross-section can be compacted in the following
way:
\ba
\frac{\d\sigma}{\d\ct} &= \frac{\d\sigma}{\d\ct}\Big|_{\mathrm{Born}}
+ \frac{\d\sigma}{\d\ct} \Big|_{\mathrm{1-Loop}}
\label{dcscost} \\
&= \frac{\d\sigma}{\d\ct}\Big|_{\mathrm{Born}} +
  \frac 1{32\pi} \frac{\beta_f}{\beta_e} C_f 2\re \left( \sum_{i=1}^4
    {F^{(0)}_i}^*\tilde{F}^{(1)}_i \right) \,,
\label{dcscompact}
\ea
with form factors $\tilde{F}_i^{(1)}$ given by
\be
\tilde{F}^{(1)}_i \equiv  \frac{1}{s}\sum_{j,k=1}^{4} M_{1,i}^\dagger
M_{j,k}^{\phantom{\dagger}} F^{(1)}_{j,k} \,,
\label{defftilde}
\ee
 that include the corresponding 
kinematical terms from the product of matrix elements\footnote{%
  Since the corrections are of \oa with respect to the Born 
  cross-section, we neglected the effect of the electron mass here.} 
together with the one-loop form factors $F_{j,k}^{(1)}$, carefully 
defined in \cite{Fleischer:2003kk} and corresponding to the basis 
(\ref{fullbasis}).  The explicit expressions for these form factors $\tilde{F}_i^{(1)}$ are:
\ba
\tilde{F}^{(1)}_1
\equiv &+4 m_f^2 \left( F^{(1)}_{1,1} - F^{(1)}_{3,1} m_f \right) \nonumber\\
& + s \bigg\{ F^{(1)}_{1,1}  + \big( F^{(1)}_{3,1} - 2F^{(1)}_{4,1} +
2F^{(1)}_{4,4} \big)m_f -F^{(1)}_{2,4}m_f^2 
\nonumber\\
& \phantom{+s \bigg\{}+ \beta_f\ct  \left( 2F^{(1)}_{1,4} - F^{(1)}_{2,1}m_f^2 \right) +
  \beta_f^2\cos^2\theta \left( F^{(1)}_{1,1} - F^{(1)}_{3,1}m_f \right) \bigg\} \nonumber\\
&+ \frac{s^2}{4} (1-\beta_f^2\cos^2\theta) \left( F^{(1)}_{2,4} +  \beta_f\ct~F^{(1)}_{2,1} \right)
\label{fbar1} 
\\
\tilde{F}^{(1)}_2
\equiv & - 4F^{(1)}_{1,2}m_f^2 \nonumber\\
& + s \bigg\{  F^{(1)}_{1,2} - F^{(1)}_{2,3}m_f^2 
  + \beta_f\ct \left( 2F^{(1)}_{1,3}  + 2\big(F^{(1)}_{4,2} - F^{(1)}_{4,3}\big)m_f -F^{(1)}_{2,2}m_f^2 \right)
\nonumber\\
& \phantom{+s \bigg\{} + \beta_f^2\cos^2\theta F^{(1)}_{1,2} \bigg\}
+ \frac{s^2}{4} (1-\beta_f^2\cos^2\theta) \left( F^{(1)}_{2,3} + \beta_f\cos\theta~F^{(1)}_{2,2} \right)
\label{fbar2}
\\
\tilde{F}^{(1)}_3
\equiv & + 4m_f^2 \left(F^{(1)}_{1,3} - F^{(1)}_{3,3}m_f\right)\nonumber\\
& + s \bigg\{ F^{(1)}_{1,3}  + \big(F^{(1)}_{3,3} + 2F^{(1)}_{4,2} - 2F^{(1)}_{4,3} \big) m_f -  F^{(1)}_{2,2}m_f^2 \nonumber\\
& \phantom{+s \bigg\}} + \beta_f\ct \left( + 2F^{(1)}_{1,2} -  F^{(1)}_{2,3}m_f^2 \right) 
  + \beta_f^2\cos^2\theta  \left( F^{(1)}_{1,3} - F^{(1)}_{3,3}m_f \right) \bigg\}\nonumber\\
&+ \frac{s^2}{4}(1-\beta_f^2\cos^2\theta) \left( F^{(1)}_{2,2} + \beta_f\ct~F^{(1)}_{2,3} \right)
\label{fbar3} 
\\
\tilde{F}^{(1)}_4
\equiv & - 4F^{(1)}_{1,4}m_f^2\nonumber\\
& + s \bigg\{ F^{(1)}_{1,4} - F^{(1)}_{2,1}m_f^2
  + \beta_f\ct \left( 2F^{(1)}_{1,1} + 2 \big(- F^{(1)}_{4,1} + F^{(1)}_{4,4}\big)m_f - F^{(1)}_{2,4}m_f^2 \right)
\nonumber\\
&  \phantom{+s \bigg\}}   + \beta_f^2\cos^2\theta F^{(1)}_{1,4} \bigg\}
+\frac{s^2}{4}(1 - \beta_f^2\cos^2\theta) \left( F^{(1)}_{2,1} +
  \beta_f\ct~F^{(1)}_{2,4} \right) \,.
\label{fbar4}
\ea

Many technical details of the underlying calculations have been
described in \cite{Fleischer:2003kk,Beenakker:1991ca}.


\section{Numerical results}

In this section we present the numerical results for various final
states at two typical LC energies: 500 GeV and 1 TeV. We performed two
fixed-order calculations, i.e.\ no higher-order corrections such as
photon exponentiation have been taken into account.  The MPI Munich
group performed a fully automated calculation using \FA\
\cite{Kublbeck:1990xc,Hahn:2000kx} and \FC\ \cite{Hahn:1998yk}, where
the fermionic structures were evaluated in the Weyl--van-der-Waerden
formalism \cite{Hahn:2002vc} rather than by introducing helicity matrix
elements $M_{j,k}$ as outlined before. The numbers of the Zeuthen/CERN
group are obtained from a partly automated calculation with \diana\
\cite{Tentyukov:1999is} and \form\ \cite{FORM,Vermaseren:2000nd}, using
a \fortran\ code obtainable from \cite{FRWL:2003aa}. Both codes use
\looptools\ \cite{Hahn:1998yk}.

We assume the same input values as were used in
\cite{Fleischer:2002rn,Fleischer:2002nn,Fleischer:2003kk}.  They are
described in \tab{IPS}.

\begin{table}[H]
\begin{center}
\begin{tabular}{|c|c|}
\hline
Fermion Masses & Boson Masses \& Widths\\
\hline
\begin{math}
\begin{array}{l@{=~}r@{.}l@{\mathrm{~GeV}}}
m_{\nu}&0&0\\
m_e&0&00051099907\\
m_{\mu}&0&105658389\\
m_{\tau}&1&77705\\
m_u&0&062\\
m_c&1&5\\
m_t&173&8\\
m_d&0&083\\
m_s&0&215\\
m_b&4&7
\end{array}
\end{math}
&
\begin{math}
\begin{array}{l@{=~}r@{.}l@{\mathrm{~GeV}}}
m_{\gamma}&0&0\\
m_W&80&4514958\\
m_Z&91&1867\\
m_H&120&0\\
\Gamma_W&0&0\\
\Gamma_Z&0&0\\
\Gamma_H&0&0
\end{array}
\end{math}
\\
\hline
\hline
\multicolumn{2}{|c|}{Other Parameters}\\
\hline
\multicolumn{2}{|c|}
{
\begin{math}
\begin{array}{l@{=~}l}
\alpha&1/137.03599976\\
E^{\mathrm{max}}_{\gamma_\mathrm{soft}}& \phantom{\Big|}\sqrt{s}/10\\
(\hbar c)^2& 0.38937966 \cdot 10^9 \mathrm{GeV}^2 \text{pb}
\end{array}
\end{math}
}\\
\hline
\end{tabular}
\caption{Input parameter set.}
\label{IPS}
\end{center}
\end{table}

The cross-sections shown below depend on the maximum soft-photon energy
$E^{\mathrm{max}}_{\gamma_\mathrm{soft}}$.  This dependence should
eventually cancel when hard-photon radiation is added, but only for
sufficiently
small values of $E^{\mathrm{max}}_{\gamma_\mathrm{soft}}$.  The value
$E^{\mathrm{max}}_{\gamma_\mathrm{soft}} = \sqrt{s}/10$, which was used
in the numerical evaluation, is by far too large if one aims at a high
numerical accuracy after combination with real, hard-photon emission. It
has been chosen here nevertheless because it ensures positive
cross-section values of a realistic order of magnitude. Even for this
large value, however, the numerical change in the combined soft- and
hard-photon corrections compared to more realistic values of
$E^{\mathrm{max}}_{\gamma_\mathrm{soft}}$ is at most few per cent at
$\sqrt{s} = 500$ GeV and few per mill at $\sqrt{s} = 1$ TeV
\cite{Fleischer:2002nn}. 


The following differential cross-sections are compared:
\begin{itemize}
\item $\frac{\d\sigma}{\d\ct}\Big|_{\mathrm{Born}}$:
  Born cross-section

\item $\frac{\d\sigma}{\d\ct}\Big|_{\mathrm{B+weak}}$:
  Interference of Born with one-loop virtual weak corrections.  The 
  running of the electromagnetic coupling is also included in the
  tables\footnote{%
        This is not the case for the plots, where the running of the
        electromagnetic coupling is not included into  
        the weak contributions.}

\item $\frac{\d\sigma}{\d\ct}\Big|_{\mathrm{B+w+QED+soft}}$:
  The QED + soft photon emission (with
  \mbox{$E_{\gamma_{\mathrm{soft}}}^{\mathrm{max}}=\sqrt{s}/10$)} is 
  added to the previous contributions
\end{itemize}

The main numerical results are documented in
Tabs.~\ref{table_tau_500}--\ref{table_t_1000}. Compared to
\cite{Fleischer:2002rn}, the agreement between our calculations for
top-pair production has been improved by a factor $10^3$.  This has been
achieved thanks to a closer contact between both groups and a more
methodological programming in the \fortran\ code \topfit.  The agreement
reaches now 11 digits of technical precision, for all flavours studied.

\begin{figure}[H]
\begin{tabular}{cc}
\vspace{-0.5cm}
a) {\large $\tau$} production. & b) {\large $b$} production.\\
\vspace{+0.5cm}
 \includegraphics[scale=0.5]{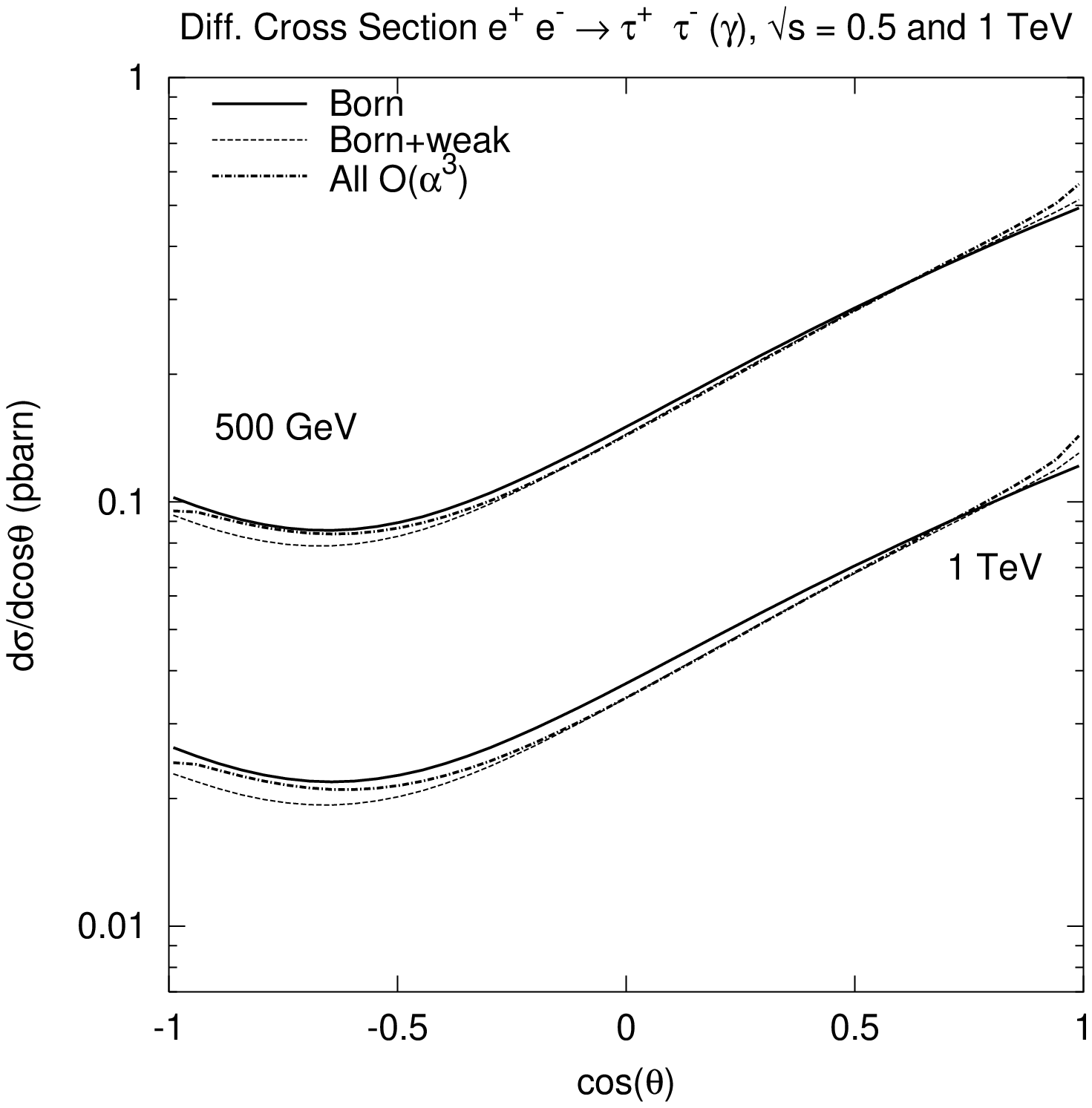} &
 \includegraphics[scale=0.5]{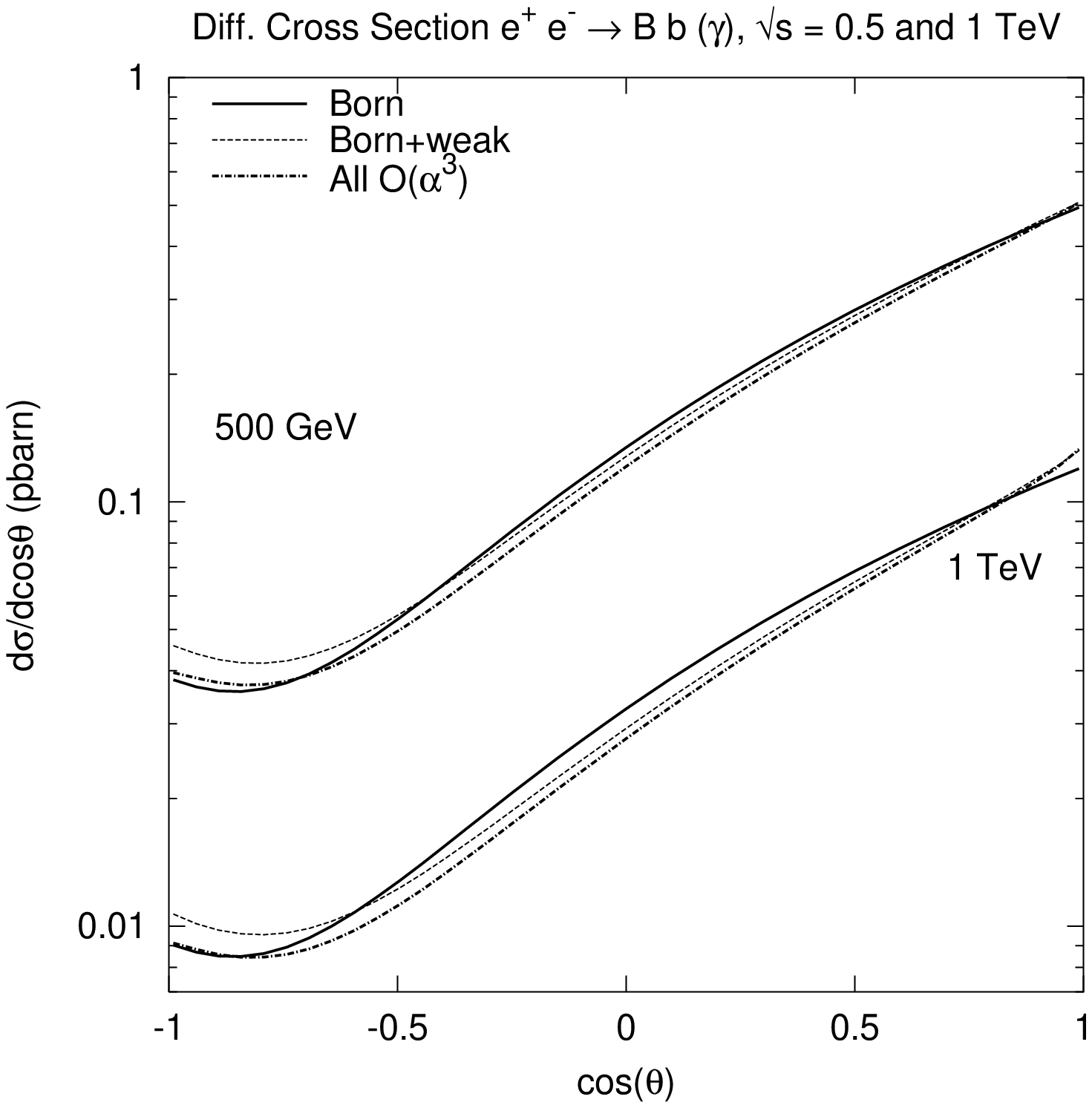} \\
\vspace{-0.5cm}
c) {\large $c$} production. & d) {\large $t$} production.\\
 \includegraphics[scale=0.5]{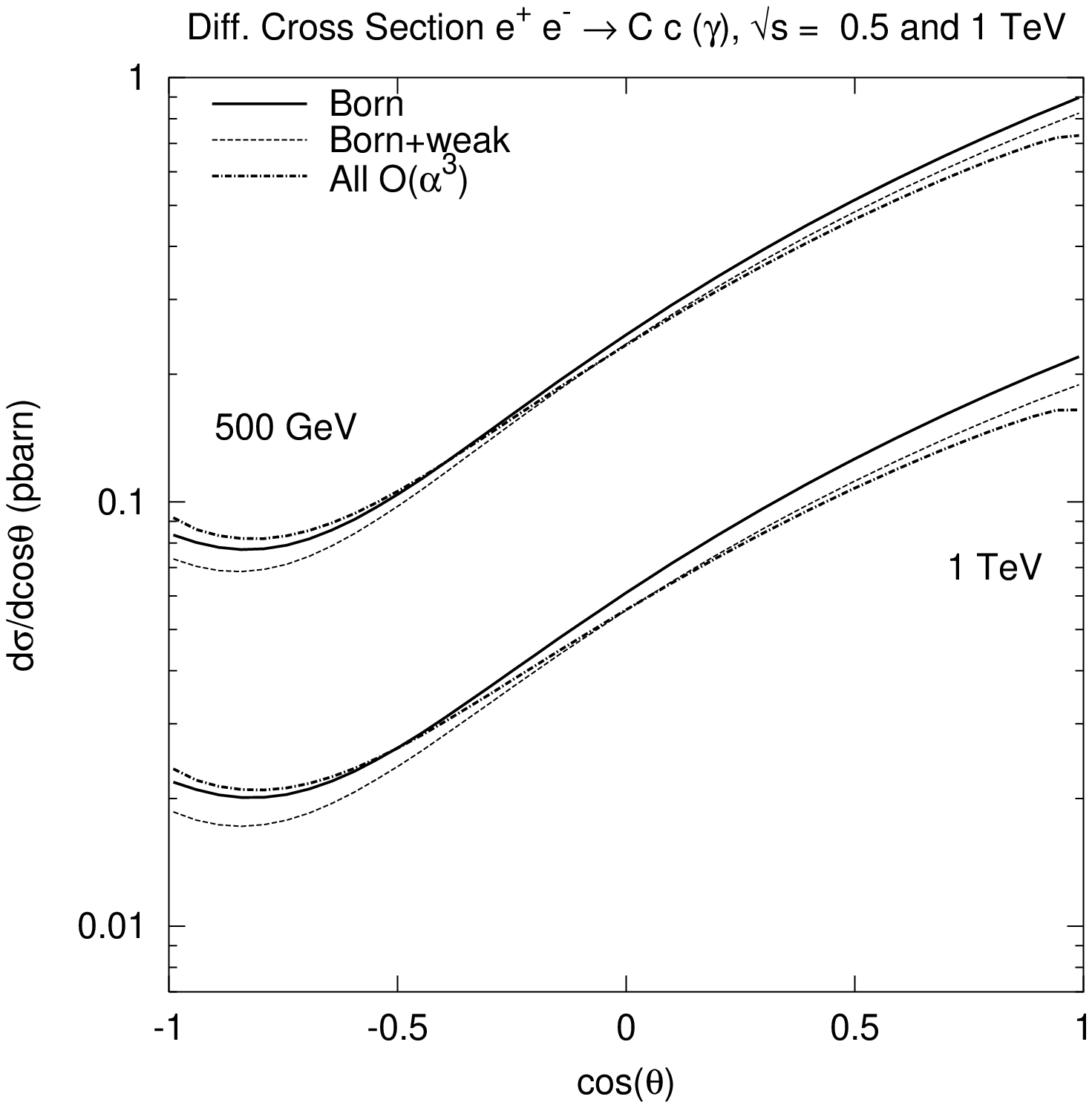} &
 \includegraphics[scale=0.5]{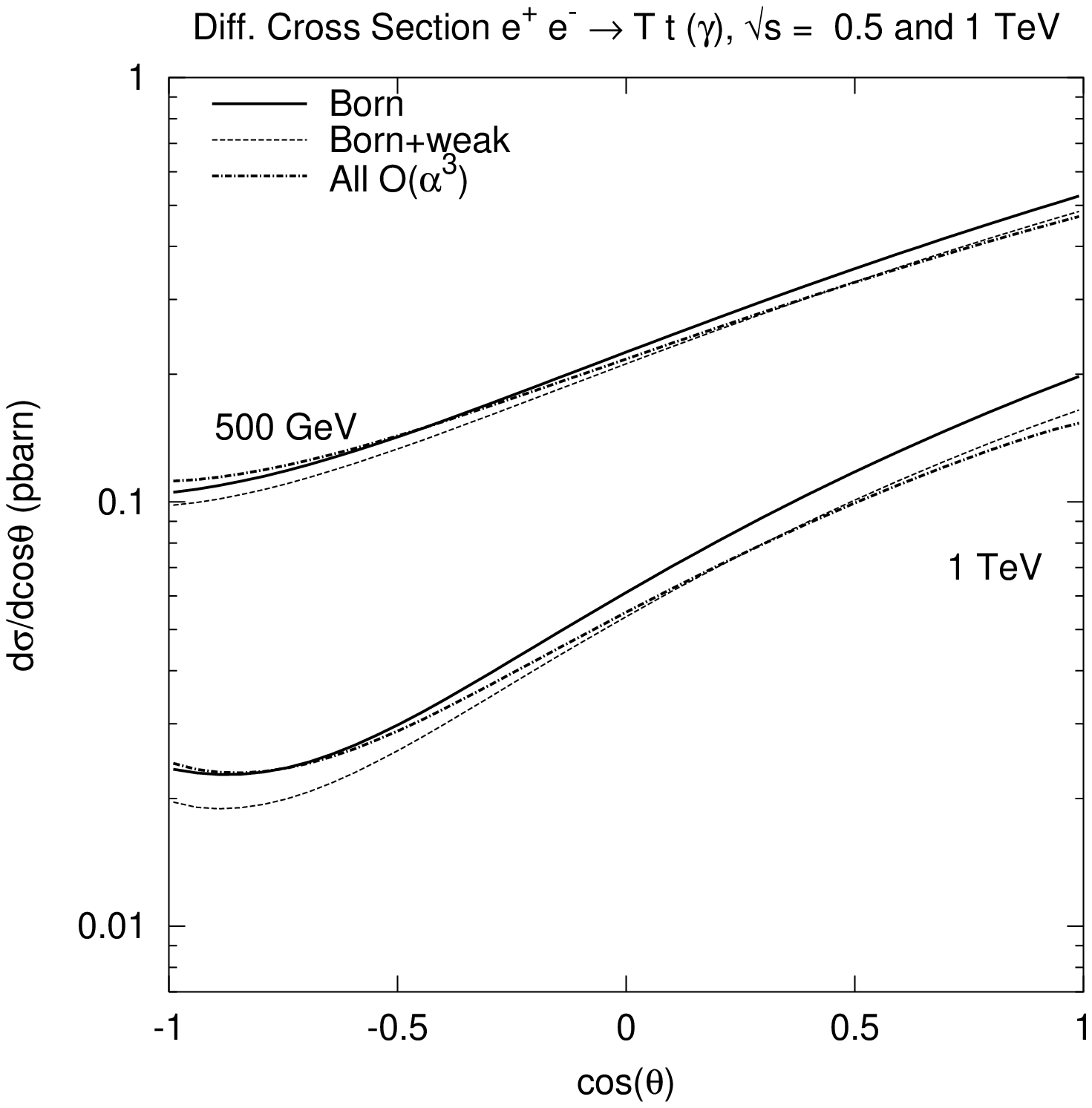} \\
\end{tabular}
\caption{Comparison of differential cross-sections. Solid line
  stands for Born, dashed for Born+weak (without running coupling), and
dashed-dotted for complete $\mathcal{O}(\alpha)$ (i.e. Born+weak+QED+soft).}
\label{dcs_taubct}
\end{figure}

Finally, in Fig.~\ref{dcs_taubct}, we give an overview of the
differential cross-sections for the different flavours at two typical
collider energies.


\newpage

\def\begintab#1#2{%
  \begin{array}{|r|l|l|l|l|}
  \hline
  \multicolumn{5}{|c|}{\vphantom{\Big|}e^+e^-\to #1\qquad \sqrt s = \text{#2}} \\
  \hline\vphantom{\Big|}
  \ct &
  \left[\frac{\d\sigma}{\d\ct}\right]_{\textrm{Born}}/\textrm{pb} &
  \left[\frac{\d\sigma}{\d\ct}\right]_{\textrm{B+weak}}/\textrm{pb} &
  \left[\frac{\d\sigma}{\d\ct}\right]_{\textrm{B+w+QED+soft}}/\textrm{pb} &
  \text{Program} \\
  \hline\hline
}

\begin{table}[H]
{\footnotesize
$$
\begintab{\tau^+\tau^-}{500 GeV}
-0.9 & 0.94591~02171~8632{\red9} \emo & 0.10860~60371~9{\red2303} & 0.92419~02671~1{\red4061} \emo & \textsc{Topfit}\\
-0.9 & 0.94591~02171~8632{\red7} \emo & 0.10860~60371~9{\red3233} & 0.92419~02671~1{\red8656} \emo & \mathrm{FA/FC}\\
\hline
-0.5 & 0.89298~53117~7985{\red8} \emo & 0.10025~68354~16{\red001} & 0.86699~48248~6{\red5248} \emo & \textsc{Topfit}\\
-0.5 & 0.89298~53117~7985{\red6} \emo & 0.10025~68354~16{\red428} & 0.86699~48248~6{\red9477} \emo & \mathrm{FA/FC}\\
\hline
 0.0 & 0.15032~16827~75192 & 0.16418~09556~08{\red258} & 0.14359~79492~086{\red48} & \textsc{Topfit}\\
 0.0 & 0.15032~16827~75192 & 0.16418~09556~07{\red903} & 0.14359~79492~086{\red18} & \mathrm{FA/FC}\\
\hline
 0.5 & 0.28649~90174~53525 & 0.31504~05045~0{\red7441} & 0.28258~86777~59{\red811} & \textsc{Topfit}\\
 0.5 & 0.28649~90174~53525 & 0.31504~05045~0{\red6135} & 0.28258~86777~59{\red161} & \mathrm{FA/FC}\\
\hline
 0.9 & 0.44955~18970~14604 & 0.50904~21673~7{\red8790} & 0.47648~29191~20{\red038} & \textsc{Topfit}\\
 0.9 & 0.44955~18970~14604 & 0.50904~21673~7{\red6612} & 0.47648~29191~19{\red623} & \mathrm{FA/FC}\\
\hline
\end{array}
$$
}
\caption{Differential cross-sections for selected scattering angles
 for $\tau$-production at $\sqrt s = 500$ GeV.  The three columns contain 
 the Born cross-section, Born including only the weak $\Oa$ corrections, 
 and Born including the weak and photonic $\Oa$ corrections.  For each 
 angle, the first row represents the \topfit\ result of the Zeuthen group 
 while the second stands for the \FA/\FC\ calculation of the Munich 
 group.}
\label{table_tau_500}
\end{table}

\begin{table}[H]
{\footnotesize
$$
\begintab{\tau^+\tau^-}{1 TeV}
-0.9 & 0.24337~58691~13477 \emo & 0.27641~21664~5{\red8412} \emo & 0.23440~03881~6{\red8909} \emo & \textsc{Topfit}\\
-0.9 & 0.24337~58691~13477 \emo & 0.27641~21664~6{\red0671} \emo & 0.23440~03881~7{\red0852} \emo & \mathrm{FA/FC}\\
\hline
-0.5 & 0.22648~34522~34421 \emo & 0.25087~88401~1{\red1477} \emo & 0.21435~50246~9{\red2009} \emo & \textsc{Topfit}\\
-0.5 & 0.22648~34522~34421 \emo & 0.25087~88401~1{\red2536} \emo & 0.21435~50246~9{\red3075} \emo & \mathrm{FA/FC}\\
\hline
 0.0 & 0.37338~94309~20687 \emo & 0.40075~04507~0{\red3072} \emo & 0.34538~81564~13{\red972} \emo & \textsc{Topfit}\\
 0.0 & 0.37338~94309~20687 \emo & 0.40075~04507~0{\red2276} \emo & 0.34538~81564~13{\red421} \emo & \mathrm{FA/FC}\\
\hline
 0.5 & 0.70698~59649~2371{\red5} \emo & 0.76863~25654~0{\red9100} \emo & 0.68181~23407~8{\red1333} \emo & \textsc{Topfit}\\
 0.5 & 0.70698~59649~2371{\red4} \emo & 0.76863~25654~0{\red6057} \emo & 0.68181~23407~7{\red8805} \emo & \mathrm{FA/FC}\\
\hline
 0.9 & 0.11082~80391~95421 & 0.12645~00486~28{\red998} & 0.11773~76209~15{\red053} & \textsc{Topfit}\\
 0.9 & 0.11082~80391~95421 & 0.12645~00486~28{\red487} & 0.11773~76209~14{\red679} & \mathrm{FA/FC}\\
\hline
\end{array}
$$
}
\caption{The same as \tab{table_tau_500} for $\sqrt s = 1$ TeV.}
\label{table_tau_1000}
\end{table}

\begin{table}[H]
{\footnotesize
$$
\begintab{b\bar b}{500 GeV}
-0.9 & 0.35947~21020~03927 \emo & 0.42347~36269~5{\red6878} \emo & 0.37629~38061~{\red 51582} \emo & \textsc{Topfit}\\
-0.9 & 0.35947~21020~03927 \emo & 0.42347~36269~5{\red0374} \emo & 0.37629~38061~{\red 44883} \emo & \mathrm{FA/FC}\\
\hline
-0.5 & 0.52846~99142~9459{\red 5} \emo & 0.55564~40895~{\red92051} \emo & 0.49542~16119~6{\red 4096} \emo & \textsc{Topfit}\\
-0.5 & 0.52846~99142~9459{\red 4} \emo & 0.55564~40895~{\red84646} \emo & 0.49542~16119~5{\red 7136} \emo & \mathrm{FA/FC}\\
\hline
 0.0 & 0.13444~84372~56821 & 0.13513~90019~9{\red9522} & 0.12117~62087~0{\red 2347} & \textsc{Topfit}\\
 0.0 & 0.13444~84372~56821 & 0.13513~90019~9{\red7996} & 0.12117~62087~0{\red 0907} & \mathrm{FA/FC}\\
\hline
 0.5 & 0.28324~62378~51991 & 0.29122~72277~5{\red3244} & 0.26454~12363~9{\red 5596} & \textsc{Topfit}\\
 0.5 & 0.28324~62378~51991 & 0.29122~72277~5{\red0185} & 0.26454~12363~9{\red 2671} & \mathrm{FA/FC}\\
\hline
 0.9 & 0.45066~58537~60950 & 0.48256~44834~8{\red 5869} & 0.44708~31668~1{\red 9343} & \textsc{Topfit}\\
 0.9 & 0.45066~58537~60950 & 0.48256~44834~8{\red 1057} & 0.44708~31668~1{\red 5091}
 & \mathrm{FA/FC}\\
\hline
\end{array}
$$
}
\caption{The same as \tab{table_tau_500} for $b$-production at $\sqrt s = 500$ GeV.}
\label{table_b_500}
\end{table}

\begin{table}[H]
{\footnotesize
$$
\begintab{b\bar b}{1 TeV}
-0.9 & 0.85256~94949~3876{\red9} \emt & 0.98313~19956~{\red72613} \emt & 0.86113~09362~{\red51944} \emt & \textsc{Topfit}\\
-0.9 & 0.85256~94949~3876{\red8} \emt & 0.98313~19956~{\red58270} \emt & 0.86113~09362~{\red37511} \emt & \mathrm{FA/FC}\\
\hline
-0.5 & 0.12689~55586~65297 \emo & 0.12711~32506~7{\red1243} \emo & 0.11163~82185~0{\red3862} \emo & \textsc{Topfit}\\
-0.5 & 0.12689~55586~65297 \emo & 0.12711~32506~6{\red9579} \emo & 0.11163~82185~0{\red2235} \emo & \mathrm{FA/FC}\\
\hline
 0.0 & 0.32532~44660~7607{\red3} \emo & 0.31258~65157~5{\red5267} \emo & 0.27674~41895~0{\red3390} \emo & \textsc{Topfit}\\
 0.0 & 0.32532~44660~7607{\red2} \emo & 0.31258~65157~5{\red1750} \emo & 0.27674~41894~9{\red9947} \emo & \mathrm{FA/FC}\\
\hline
 0.5 & 0.68639~85356~49626 \emo & 0.69302~03325~8{\red9997} \emo & 0.62501~12097~1{\red4961} \emo & \textsc{Topfit}\\
 0.5 & 0.68639~85356~49626 \emo & 0.69302~03325~8{\red2884} \emo & 0.62501~12097~0{\red7973} \emo & \mathrm{FA/FC}\\
\hline
 0.9 & 0.10923~62308~06567 & 0.12127~77274~4{\red8650} & 0.11240~86957~3{\red9236} & \textsc{Topfit}\\
 0.9 & 0.10923~62308~06567 & 0.12127~77274~4{\red7528}& 0.11240~86957~3{\red8153} & \mathrm{FA/FC}\\
\hline
\end{array}
$$
}
\caption{The same as \tab{table_tau_500} for $b$-production at $\sqrt s = 1$ TeV.}
\label{table_b_1000}
\end{table}

\begin{table}[H]
{\footnotesize
$$
\begintab{c\bar c}{500 GeV}
-0.9 & 0.78403~69156~96992 \emo & 0.91244~84607~{\red87569} \emo & 0.83668~39315~{\red90920} \emo & \textsc{Topfit}\\
-0.9 & 0.78403~69156~96992 \emo & 0.91244~84607~{\red99371} \emo & 0.83668~39316~{\red04269} \emo & \mathrm{FA/FC}\\
\hline
-0.5 & 0.10411~12875~82399 & 0.11650~15689~39{\red071} & 0.10590~20427~16{\red561} & \textsc{Topfit}\\
-0.5 & 0.10411~12875~82399 & 0.11650~15689~39{\red412} & 0.10590~20427~16{\red692} & \mathrm{FA/FC}\\
\hline
 0.0 & 0.24770~82888~4590{\red1} & 0.26255~80017~6{\red8786} & 0.23448~15990~2{\red5778} & \textsc{Topfit}\\
 0.0 & 0.24770~82888~4590{\red0} & 0.26255~80017~6{\red7528} & 0.23448~15990~2{\red3961} & \mathrm{FA/FC}\\
\hline
 0.5 & 0.51515~25192~73431 & 0.53094~95526~1{\red9036} & 0.46371~41775~1{\red7198} & \textsc{Topfit}\\
 0.5 & 0.51515~25192~73431 & 0.53094~95526~1{\red5566} & 0.46371~41775~1{\red2847} & \mathrm{FA/FC}\\
\hline
 0.9 & 0.81827~79086~1355{\red7} & 0.83043~43356~6{\red1887} & 0.70026~97050~2{\red9472} & \textsc{Topfit}\\
 0.9 & 0.81827~79086~1355{\red6} & 0.83043~43356~5{\red6199} & 0.70026~97050~2{\red1870} & \mathrm{FA/FC}\\
\hline
\end{array}
$$
}
\caption{The same as \tab{table_tau_500} for $c$-production at $\sqrt s = 500$ GeV.}
\label{table_c_500}
\end{table}

\begin{table}[H]
{\footnotesize
$$
\begintab{c\bar c}{1 TeV}
-0.9 & 0.20476~82671~10479 \emo & 0.23804~15350~7{\red4367} \emo & 0.21460~20354~0{\red3294} \emo & \textsc{Topfit}\\
-0.9 & 0.20476~82671~10479 \emo & 0.23804~15350~7{\red7280} \emo & 0.21460~20354~0{\red6337} \emo & \mathrm{FA/FC}\\
\hline
-0.5 & 0.26302~86046~48394 \emo & 0.29192~27449~2{\red8377} \emo & 0.26283~52825~1{\red9898} \emo & \textsc{Topfit}\\
-0.5 & 0.26302~86046~48394 \emo & 0.29192~27449~2{\red9292} \emo & 0.26283~52825~2{\red0679} \emo & \mathrm{FA/FC}\\
\hline
 0.0 & 0.61063~66375~8392{\red1} \emo & 0.63092~30352~2{\red7478} \emo & 0.55698~44755~9{\red1055} \emo & \textsc{Topfit}\\
 0.0 & 0.61063~66375~8392{\red0} \emo & 0.63092~30352~2{\red4633} \emo & 0.55698~44755~8{\red7819} \emo & \mathrm{FA/FC}\\
\hline
 0.5 & 0.12635~58682~75626 & 0.12548~22393~89{\red320} & 0.10778~82066~27{\red453} & \textsc{Topfit}\\
 0.5 & 0.12635~58682~75626 & 0.12548~22393~88{\red519} & 0.10778~82066~26{\red582} & \mathrm{FA/FC}\\
\hline
 0.9 & 0.20057~22407~70464 & 0.19463~36446~4{\red8183} & 0.16019~87823~3{\red2139} & \textsc{Topfit}\\
 0.9 & 0.20057~22407~70464 & 0.19463~36446~4{\red6866} & 0.16019~87823~3{\red0647} & \mathrm{FA/FC}\\
\hline
\end{array}
$$
}
\caption{The same as \tab{table_tau_500} for $c$-production at $\sqrt s = 1$ TeV.}
\label{table_c_1000}
\end{table}

\begin{table}[H]
{\footnotesize
$$
\begintab{t\bar t}{500 GeV}
-0.9 & 0.10883~91940~76039 & 0.12425~90371~32{\red943} & 0.11408~40955~7{\red7861} & \textsc{Topfit}\\
-0.9 & 0.10883~91940~76039 & 0.12425~90371~33{\red664} & 0.11408~40955~7{\red8964} & \mathrm{FA/FC}\\
\hline
-0.5 & 0.14227~50693~93371 & 0.15684~83718~76{\red069} & 0.14308~12051~655{\red11} & \textsc{Topfit}\\
-0.5 & 0.14227~50693~93371 & 0.15684~83718~76{\red250} & 0.14308~12051~655{\red81} & \mathrm{FA/FC}\\
\hline
 0.0 & 0.22547~04640~33559 & 0.24026~68040~30{\red724} & 0.21718~80097~6{\red7412} & \textsc{Topfit}\\
 0.0 & 0.22547~04640~33559 & 0.24026~68040~30{\red032} & 0.21718~80097~6{\red6323} & \mathrm{FA/FC}\\
\hline
 0.5 & 0.35466~64703~33217 & 0.36888~65069~9{\red4389} & 0.32933~72739~5{\red1692} & \textsc{Topfit}\\
 0.5 & 0.35466~64703~33217 & 0.36888~65069~9{\red2599} & 0.32933~72739~4{\red9095} & \mathrm{FA/FC}\\
\hline
 0.9 & 0.49114~37157~67761 & 0.50333~75116~0{\red5520} & 0.44290~81673~5{\red1494} & \textsc{Topfit}\\
 0.9 & 0.49114~37157~67761 & 0.50333~75116~0{\red2681} & 0.44290~81673~4{\red6094} & \mathrm{FA/FC}\\
\hline
\end{array}
$$
}
\caption{The same as \tab{table_tau_500} for $t$-production at $\sqrt s = 500$ GeV.}
\label{table_t_500}
\end{table}

\begin{table}[H]
{\footnotesize
$$
\begintab{t\bar t}{1 TeV}
-0.9 & 0.22785~42327~32090 \emo & 0.25521~28532~9{\red8051} \emo & 0.23101~70508~0{\red5040} \emo & \textsc{Topfit}\\
-0.9 & 0.22785~42327~32090 \emo & 0.25521~28533~0{\red0748} \emo & 0.23101~70508~0{\red7714} \emo & \mathrm{FA/FC}\\
\hline
-0.5 & 0.29782~13110~31861 \emo & 0.31863~48943~5{\red9857} \emo & 0.28823~01902~0{\red0931} \emo & \textsc{Topfit}\\
-0.5 & 0.29782~13110~31861 \emo & 0.31863~48943~6{\red0711} \emo & 0.28823~01902~0{\red1653} \emo & \mathrm{FA/FC}\\
\hline
 0.0 & 0.61180~06742~2503{\red9} \emo & 0.61591~61295~7{\red7963} \emo & 0.54950~88904~8{\red8739} \emo & \textsc{Topfit}\\
 0.0 & 0.61180~06742~2503{\red8} \emo & 0.61591~61295~7{\red5474} \emo & 0.54950~88904~8{\red5894} \emo & \mathrm{FA/FC}\\
\hline
 0.5 & 0.11774~69498~88318 & 0.11404~76860~51{\red226} & 0.99417~00898~3{\red9905} \emo & \textsc{Topfit}\\
 0.5 & 0.11774~69498~88318 & 0.11404~76860~50{\red527} & 0.99417~00898~3{\red2292} \emo & \mathrm{FA/FC}\\
\hline
 0.9 & 0.18112~20970~86446 & 0.17134~61927~2{\red2790} & 0.14426~23325~4{\red1248} & \textsc{Topfit}\\
 0.9 & 0.18112~20970~86446 & 0.17134~61927~2{\red1645} & 0.14426~23325~4{\red0061} & \mathrm{FA/FC}\\
\hline
\end{array}
$$
}
\caption{The same as \tab{table_tau_500} for $t$-production at $\sqrt s = 1$ TeV.}
\label{table_t_1000}
\end{table}



\newpage

\providecommand{\href}[2]{#2}

\end{document}